\documentclass[conference]{IEEEtran}
\usepackage{array}
\usepackage{adjustbox}
\usepackage{amsmath}
\usepackage{fancyhdr}
\usepackage{graphicx}
\usepackage{float}
\usepackage[LGRgreek]{mathastext}
\usepackage{lipsum}
\usepackage{multirow}
\usepackage{subfigure}
\usepackage{url}
\usepackage{varwidth}
\usepackage{booktabs}
\usepackage[table]{xcolor}
\usepackage{tabularx}
\usepackage{arydshln}
\usepackage{datatool}
\usepackage{courier}
\usepackage{lscape}
\usepackage{pdflscape}
\usepackage{siunitx}

\DeclareSIUnit\px{px}
\usepackage{bbding}
\usepackage{pifont}
\usepackage{wasysym}
\usepackage{amssymb}
\IEEEoverridecommandlockouts   
\renewcommand\IEEEkeywordsname{Keywords}

\usepackage{acro}
\DeclareAcronym{ac}{
  short = AC,
  long  = Auto-encoder
}

\DeclareAcronym{cad}{
  short = CAD,
  long  = Computer Aided Diagnosis
}

\DeclareAcronym{gt}{
  short = GT,
  long  = Ground Truth
}

\DeclareAcronym{acr}{
  short = ACR,
  long  = American College of Radiology
}

\DeclareAcronym{fa}{
  short = FA,
  long  = Fibro-Adenoma
}

\DeclareAcronym{dic}{
  short = DIC,
  long  = Ductal Inflating Carcinoma
}

\DeclareAcronym{ilc}{
  short = ILC,
  long  = Inflating Lobular Carcinoma
}

\DeclareAcronym{dr}{
  short = DR,
  long  = diabetic retinopathy
}

\DeclareAcronym{dme}{
  short = DME,
  long  = diabetic macular edema
}

\DeclareAcronym{oct}{
  short = OCT,
  long  = optical coherence tomography
}

\DeclareAcronym{sdoct}{
  short = SD-OCT,
  long  = Spectral Domain OCT
}

\DeclareAcronym{amd}{
  short = AMD,
  long = age-related macular degeneration
}

\DeclareAcronym{pca}{
  short = PCA,
  long = Principal Component Analysis
}

\DeclareAcronym{nlm}{
  short = NLM,
  long = Non-Local Means
}

\DeclareAcronym{who}{
  short = WHO,
  long = World Health Organization
}

\DeclareAcronym{ncd}{
  short = NCD,
  long = Non-Communicating Diseases
}

\DeclareAcronym{ria}{
  short = RIA,
  long = Retinopathy Image Analysis
}

\DeclareAcronym{bm3d}{
  short = BM3D,
  long = block matching 3D filtering
}

\DeclareAcronym{snr}{
  short = SNR,
  long = Signal-to-Noise Ratio
}

\DeclareAcronym{psnr}{
  short = PSNR,
  long = Peak Signal-to-Noise Ratio
}

\DeclareAcronym{ssdc}{
  short = SSDC,
  long = Subspace-based Spatial Domain Constraint
}

\DeclareAcronym{dct}{
  short = DCT,
  long = Discrete Cosine Transform
}

\DeclareAcronym{obnlm}{
  short = OB-NLM,
  long = Optimized Bayesian NLM
}

\DeclareAcronym{pgpd}{
  short = PGPD,
  long = Patch Group Based nonlocal self-similarity prior learning for image Denoising
}

\DeclareAcronym{gmm}{
  short = GMM,
  long = Gaussian Mixture Models
}

\DeclareAcronym{db}{
  short = $dB$,
  long = decibels
}
\DeclareAcronym{svm}{
    short = SVM, 
    long = support vector machine
}
\DeclareAcronym{bow}{
    short = BoW, 
    long = bag of words
}
\DeclareAcronym{rf}{
    short = RF, 
    long = random forest
}
\DeclareAcronym{auc}{
    short = AUC, 
    long = Area Under the Curve
}
\DeclareAcronym{rbf}{
    short = RBF, 
    long = radial basis function
}
\DeclareAcronym{lbp}{
    short = LBP, 
    long = local binary pattern
}
\DeclareAcronym{lbptop}{
    short = LBP-TOP,
    long = Three Orthogonal Planes
}
\DeclareAcronym{hog}{
    short = HoG, 
    long = histogram of oriented gradients
}
\DeclareAcronym{se}{
    short = SE, 
    long = sensitivity
}
\DeclareAcronym{sp}{
    short = SP, 
    long = specificity
}
\DeclareAcronym{acc}{
    short = ACC, 
    long = accuracy
}
\DeclareAcronym{pre}{
    short = PRE, 
    long = precision
}
\DeclareAcronym{f1s}{
    short = F1-s, 
    long = F1-score
}
\DeclareAcronym{ltpocv}{
    short = LTPO-CV, 
    long = leave-two-patients-out cross-validation
}
\DeclareAcronym{seri}{
    short = SERI, 
    long = Singapore Eye Research Institute
}

\usepackage[switch,columnwise]{lineno}
\modulolinenumbers[5]

\begin{document}


\author{\IEEEauthorblockN{Khaled Al-Saih$^{1}$$^{*}$}
\IEEEauthorblockA{\\$^{1}$Laboratoire LIMOS, CNRS UMR 6158,\\ Université Clermont-Auvergne,\\ 
63170 Aubiere, France\\
$^{*}$E-mail: khaled.al\_saih@uca.fr}
\and
\IEEEauthorblockN{Fares Al-Shargie$^2$}
\IEEEauthorblockA{\\$^2$The School of Health Professions, \\ Department of Rehabilitation and Movement Sciences,\\ Rutgers University, Newark, NJ 07102 USA\\  E-mail: fares.yahya@rutgers.edu
}
\and
\IEEEauthorblockN{Mohammed Isam Al-hiyali$^3$}
\IEEEauthorblockA{\\$^3$Medical Instruments Technology Engineering Department, \\ AL Mansour University College,\\ Baghdad, 10068, Iraq\\  E-mail: Eng.mohammedissam@gmail.com
}
\and
\IEEEauthorblockN{Reham Alhejaili$^4$}
\IEEEauthorblockA{\\$^4$Department of Computer Science and Artificial Intelligence, \\ College of Computer Science and Engineering,\\ University of Jeddah, Jeddah 23218, Saudi Arabia\\  E-mail: ralhejaili@uj.edu.sa
}
}

%
%
%

\title{Patch-Based and Non-Patch-Based inputs Comparison into Deep Neural Models: Application for the Segmentation of Retinal Diseases on Optical Coherence Tomography Volumes}


    

\maketitle

\begin{abstract}
Worldwide, sight loss is commonly occurred by retinal diseases, with age-related macular degeneration (AMD) being a notable facet that affects elderly patients.
Approaching 170 million persons wide-ranging have been spotted with AMD, a figure anticipated to rise to 288 million by 2040.
For visualizing retinal layers, optical coherence tomography (OCT) dispenses the most compelling non-invasive method.
Frequent patient visits have increased the demand for automated analysis of retinal diseases, and deep learning networks have shown promising results in both image and pixel-level 2D scan classification.
However, when relying solely on 2D data, accuracy may be impaired, especially when localizing fluid volume diseases.
The goal of automatic techniques is to outperform humans in manually recognizing illnesses in medical data.
In order to further understand the benefit of deep learning models, we studied the effects of the input size.
The dice similarity coefficient (DSC) metric showed a human performance score of 0.71 for segmenting various retinal diseases.
Yet, the deep models surpassed human performance to establish a new era of advancement of segmenting the diseases on medical images.
However, to further improve the performance of the models, overlapping patches enhanced the performance of the deep models compared to feeding the full image.
The highest score for a patch-based model in the DSC metric was 0.88 in comparison to the score of 0.71 for the same model in non-patch-based for SRF fluid segmentation.
The objective of this article is to show a fair comparison between deep learning models in relation to the input (Patch-Based vs. NonPatch-Based).

\end{abstract}
%
\IEEEpeerreviewmaketitle
\acresetall

\section{Introduction}\label{sec:intro}
\begin{figure*}[t!]
\begin{center}
\centering

   \subfigure[Cirrus device]{\includegraphics[width=0.3\textwidth, height = 0.15\textheight]{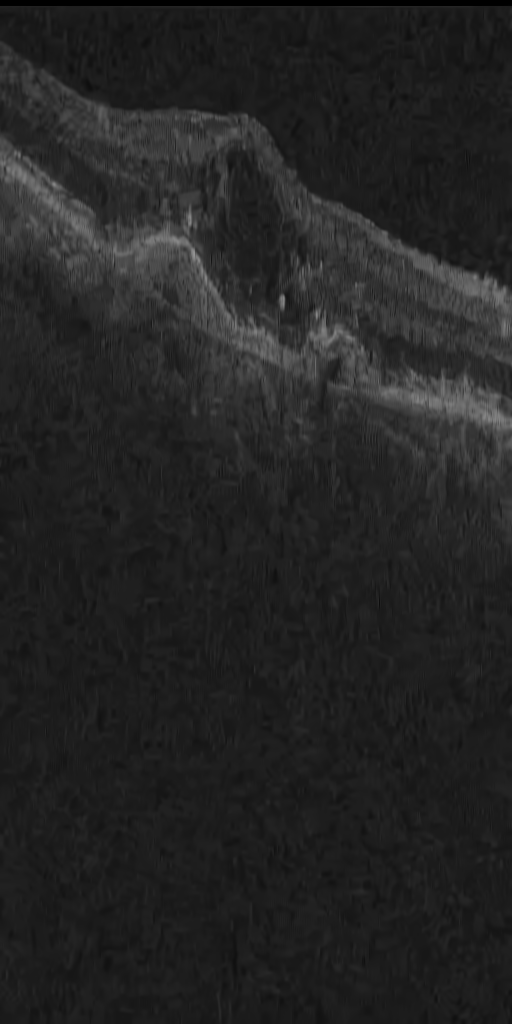}}\
   \subfigure[Spectralis device]{\includegraphics[width=0.3\textwidth,height = 0.15\textheight]{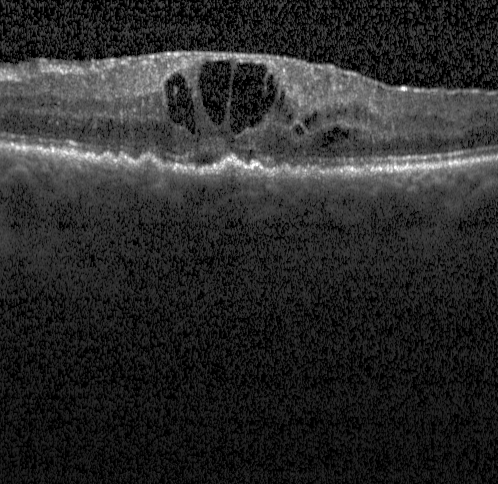}}\
   \subfigure[Topcon device]{\includegraphics[width = 0.3\textwidth,height = 0.15\textheight]{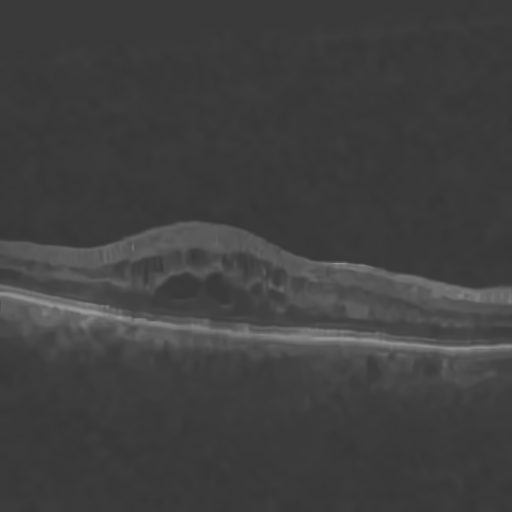}}
   \
\end{center}
    \caption{RETOUCH challenge training images from various vendors}
    \label{fig:exampleoptima}
\end{figure*}

Macular degeneration resulting from age is one of the most regular retinal diseases globally, fundamentally striking older individuals.
Prompt and constructive treatment of acute age-related macular degeneration (AMD) are crucial to prevent the loss of the eye vision.
The retinal maladies studied in this article include intraretinal fluid (IRF), subretinal fluid (SRF), and pigment epithelial detachment (PED).
AMD disorder is healed effectively using the anti-vascular endothelial growth factor (anti-VEGF) therapy \cite{fung2007optical}.
The process of handling and treating retinal malfunction requires one to couple or many doctor visits to assess the effectiveness of the anti-VEGF treatment based on the status of the eye.
Following each engagement, each patient engages in an SD-OCT scan to examine the current volume of the retinal malady, which could lead to further decision-making.

Conventionally, ophthalmologists manually segment retinal layers and fluids on an image-by-image wise, a perspective that is tedious, time-taking and susceptible to intra-rater and inter-rater variability.
Current research on automated deep models for segmenting AMD malfunction has confirmed that they can render meaningful assistance to eye specialists in diagnosing lesions.
In contrast, segmenting various fluids is a more strenuous task than segmenting layers.
The sizes, locations, and shapes of each fluid within a diseased retina vary notably within the same patient.
The dissimilarity in fluids impedes automatic segmentation.
A remarkable obstacle in the development of algorithms is the contretemps between the professional graders.
The segregation between certain fluids is not easily discernible, as illustrated by the B-scan images in Fig.~\ref{fig:exampleoptima}, which hinders the convolutional neural network (CNN) filter-imposed edge detection.
Variations in the retina's morphology are visualized through optical coherence tomography (OCT) imaging.
The OCT scanners produce and generate volumetric images.
Several well-known sellers are available, including Cirrus, Spectralis, Topcon, and Nidek.
Every device or scanner has its own distinctive implementation technicalities and differs in terms of resolution and the number of images produced per volume.
The modalities of retinal imaging, similar to other medical imaging models, are impacted by the movement of the object during the scanning progress.
The OCT devices are further impacted by movement of the eyes.

The principal purpose of this study is to compare the performance of deep neural networks when they are fed different input sizes for the same data.
In addition, the article uses the same data at three different dimensions, 2D, 2.5D and 3D, respectively.
Furthermore, it is followed by the introduction, and the paper is constructed as follows: Sect.\,\ref{sec:rw} emphasizes the state-of-the-art studies on the age-related macular degeneration (AMD) segmentation applied to the RETOUCH dataset.
Sect~\ref{sec:method} explains the deep models used as well as the input sizes and followed by Sect.\,\ref{sec:exp-res} which is dedicated to the interpretation of the findings.
Finally, this article is summed-up in Sect.\,\ref{sec:con}.

\section{Related work}\label{sec:rw}

The retina has been blessed with an intense scrutiny from researchers in modern years.
The acuteness of retina impairment may impact a person's vision.
Over the past decade, a lot of  algorithms and models have been proposed to identify and classify retinal diseases using machine learning techniques, as discussed in \cite{alsaih2017machine}.
The vast majority of the available data are commonly demanded on an exclusive basis.
Typically, private data originates from a single scanner or vendor, and the main objective is to localize either one or two fluids, as described in reference \cite{girish2018segmentation}.
Many datasets have been made available for the experimentation groups, including the OPTIMA challenge public dataset for segmenting the IRF and the RETOUCH challenge public dataset for segmenting the PED, IRF, and SRF maladies
   
The RETOUCH challenge employed the released data by multiple teams who proposed various deep and machine learning methods to identify and segment  mainly three retinal cysts.
The contributing groups had different strategies, with five teams training their networks in a scan-by-scan wise without utilizing depth details.
Initially, the group of UMN, as outlined in \cite{abdolrezadeep2019re}, localized two main retinal layers, namely, the inner limiting membrane (ILM) and retinal pigment epithelium (RPE) to narrow down the exploring area for two fluids, namely, SRF and IRF.
A CNN-based model of four convolutional layers was suggested to pinpoint the locations of the SRF and IRF cysts, while an unsupervised algorithm was used to segment and find the lesions of PED.
Then, flattening the RPE layer and determining the PED size by calculating the distance between the flattened and unflattened RPE layer.
The mean DSC scores for IRF, SRF, and PED were 0.69, 0.7, and 0.76.
The SFU team, as shown in \cite{lu2017retinal}, also explores an algorithm that localizes the layers of the retina to avoid excess background details that can negatively impact network performance.
The retinal slices were then augmented and smoothed to supply a diverse range of forms and positioning for improved learning.
The model of U-net is employed with an additional channel.
Following this, the method of random forest was utilized to remove the false-positive pixels , resulting in mean DSC scores of 0.81, 0.75, and 0.74 for the segmentation of IRF, SRF, and PED, respectively.

The spectral variation algorithm was utilized by Helios group in \cite{yadav2017generalized} for further denoising the data.
In addition, a generalized motion pattern is employed to suppress the background after resizing every B-scan, which introduces a motion to each scan.
Then, the whole set of images are fed to a customized U-net model, which is trained with cross-entropy and the mean dice score is 0.62, 0.67, and 0.66 for the segmentation of the three retinal fluids IRF, SRF, and PED, respectively.
Furthermore, Gaussian noise and rotation is applied by the RetinaAI research team \cite{apostolopoulos2017pathological2}  to the images fitted to the networks.
The dilated convolutions replaced the vanilla convolutions with a residual unit added to the U-net architecture.
The network achieved segmentation mean dice scores of 0.73, 0.67, and 0.71 for IRF, SRF, and PED, respectively.
The group of MABIC \cite{sungdeep2019re}, used 2D information.
Two parallel U-net models are used, one for the detection and another for segmentation purposes.
Maxout activations instead of ReLU operations and a dropout layer to hinder network’s overfitting.
The intensity pixel assignment method supplied by the challenge committee was changed from 0, 1, 2, and 3 for the background, and the retinal fluids (IRF, SRF, and PED, respectively), to 0-7 scale, representing the quantity of lesions presented in each B-scan.
An average DSC segmentation score of 0.77, 0.66, and 0.71 was achieved for IRF, SRF, and PED, respectively.
   
Two additional groups, NJUST team \cite{chendeepdeep2019} and the UCF team \cite{morley2017simultaneous}, didn’t utilize 2D information but 2.5 dimensional information.
Region growing algorithm was applied to find SRF retinal cysts and the intermediate fluid is found using the Faster R-CNN model by The NJUST group.
PED fluid is detected when the RPE layer is segmented.
The bilateral filter denoised each image and achieved segmentation DSC score of 0.56, 0.53, and 0.64 for IRF, SRF, and PED fluids, respectively.
The other group, UCF team, 3D smoothed and rescaled the B-scans, then also employed an encoder-decoder model named ResNet.
Besides, the images were cropped to the borders of the layers in the retina.
Moreover, the data was augmented using the myopic warping algorithm to expand the  curve quantity in the retina.
Findings for the UCF group are represented in the mean DSC of 0.69, 0.7, and 0.76, respectively for the segmentation of IRF, SRF, and PED fluid lesions.
Finally,  \cite{ruwandeep2019re} made a contribution to the challenge and utilized three dimensional information.
The RMIT group resampled the voxels to enforce uniform spacing after the extraction of three dimensional knowledge.
An adversarial network was employed to miss the post-processing phase for having a fully automated model.
The retinal images were normalized, and histogram matching was applied.
At last, retinal images were convolved in 2D mode, leading to a mean DSC score of 0.72, 0.7, and 0.69 for IRF, SRF, and PED fluid lesion segmentation, respectively.

Previously, we have studied and compared the performance of three deep encoder-decoder models between the standard convolutional neural network elements with atrous convolutional neural network elements in \cite{alsaih2020performance}.
Also, various algorithms are proposed to the RETOUCH dataset other than the competing teams like in \cite{oh2024gcn} and \cite{pavani2024robust}.
In this paper we have taken the best performing models from our previous work in 2D input \cite{deep1alsaih}, 2.5D input \cite{alsaih2020retinal1}, and used the 3D model in \cite{alsaih2020retinal2}.
In the previous work we run the models with inputting either the full image or patches except for the 3D model, we run it only for 96x96 patch size. 
Hence, in this paper, we run the same 3D model as in \cite{alsaih2020retinal2}, but in patches format with the size of 128x128 and full size of 384x384 to draw a conclusion for the researchers in which input could benefit them when they deal with OCT volumes.

\section{Methodology}\label{sec:method}
In this work, we employed three well-known deep models and made an assumption about their performance with respect to their input.
All models are trained, tested and implemented using the MATLAB programming language.
The mean time for the training phase was between 494 mins and 673 mins.
It is important to note that, in this paper, convolution-based architectures, which are trained with full image size, have the subscript (F), for example, $Deeplabv3+_F$.
Moreover, convolution-based algorithms that are trained in patch-wise are referred with the notion (P), for example, $Deeplabv3+_P$. 
The used device has 16 GB of RAM, a GEFORCE GTX 1070 Ti GPU, and an i5-core CPU.
The Cuda version used for the experimentation is 12.5 with a cuDNN version 9.6.

\subsection{Dataset}

\begin{table}[htbp]
\centering
\caption{The Description of the RETOUCH Training Dataset.}
\label{tab:datasetdes}
\begin{tabular}{cccc}
\hline
\multirow{2}{*}{Supplier}                  & \multirow{2}{*}{B-Scans Size} & \multicolumn{2}{c}{RETOUCH}               \\ \cmidrule{3-4}
                        &                  & Training Data Size           & Testing Data Size            \\ \hline
Cirrus                  & 512 x 1024 x 128 & 24                  & 14                  \\ \hline
Spectralis              & 512 x 496 x 49   & 24                  & 14                  \\ \hline
\multirow{2}{*}{Topcon} & 512 x 885 x 128  & \multirow{2}{*}{22} & \multirow{2}{*}{14} \\ \cmidrule{2-2}
                        & 512 x 650 x 128  &                     & \\ \hline                    
\end{tabular}
\end{table}
The RETOUCH challenge dataset was employed and used in this particular study \cite{bogunovic2019retouch}.
The dataset precisely targets three types of fluids: PED, SRF, and IRF.
Zero is the chosen value for the background pixels.
The obtained OCT volumes are extracted and obtained from mainly three distinct vendors: Topcon, Cirrus, and Spectralis.
Table~\ref{tab:datasetdes} presents the number and the size of the B-scans (volumes).
This research solely utilized the training set to train, validate and test the deep neural models.
The test set is made available in the absence of their ground-truth, which is why it was excluded.
\renewcommand{\arraystretch}{2}
\begin{table*}[ht!]
\caption{\textit{Exp} - The Performance of the Deep Models in Dice Score Metric}
\centering
\label{tab:my-table11}
\begin{tabular}{ccccccccccc}
\hline
\multirow{3}{*}{Input Size} & \multirow{3}{*}{Deep Model} & \multicolumn{9}{c}{Device Name}                                                                                                              \\ \cline{3-11} 
                            &                          & \multicolumn{3}{c}{Cirrus}                    & \multicolumn{3}{c}{Spectralis}                & \multicolumn{3}{c}{Topcon}                    \\ \cline{3-11} 
                            &                          & IRF           & SRF           & PED           & IRF           & SRF           & PED           & IRF           & SRF           & PED           \\ \hline
\multirow{6}{*}{2D \cite{deep1alsaih}}         & $U-net_F$                & 0.75          & 0.74          & 0.68          & 0.86          & 0.68          & 0.61          & 0.72          & 0.85          & 0.7           \\  
                            & $U-net_P$                & \textbf{0.78} & \textbf{0.84} & \textbf{0.69} & \textbf{0.91} & \textbf{0.72} & \textbf{0.63} & \textbf{0.71} & \textbf{0.83} & \textbf{0.79} \\  
                            & $Segnet_F$               & 0.72          & 0.78          & 0.63          & 0.68          & 0.77          & 0.36          & 0.44          & 0.48          & 0.56          \\  
                            & $Segnet_P$               & \textbf{0.68} & \textbf{0.82} & \textbf{0.74} & \textbf{0.66} & \textbf{0.46} & \textbf{0.73} & \textbf{0.49} & \textbf{0.53} & \textbf{0.72} \\  
                            & $Deeplabv3+_F$           & 0.68          & 0.74          & 0.81          & 0.53          & 0.48          & 0.69          & 0.41          & 0.71          & 0.68          \\  
                            & $Deeplabv3+_P$           & \textbf{0.84} & \textbf{0.88} & \textbf{0.81} & \textbf{0.88} & \textbf{0.67} & \textbf{0.83} & \textbf{0.74} & \textbf{0.88} & \textbf{0.83} \\ \hline
\multirow{6}{*}{2.5D \cite{alsaih2020retinal1}}       & $U-net_F$                & 0.72          & 0.69          & 0.75          & 0.66          & 0.62          & 0.78          & 0.68          & 0.82          & 0.74          \\ 
                            & $U-net_P$                & \textbf{0.71} & \textbf{0.75} & \textbf{0.73} & \textbf{0.75} & \textbf{0.69} & \textbf{0.82} & \textbf{0.76} & \textbf{0.88} & \textbf{0.81} \\  
                            & $Segnet_F$               & 0.76          & 0.73          & 0.7           & 0.68          & 0.7           & 0.66          & 0.55          & 0.62          & 0.5           \\  
                            & $Segnet_P$               & \textbf{0.78} & \textbf{0.77} & \textbf{0.68} & \textbf{0.74} & \textbf{0.76} & \textbf{0.7}  & \textbf{0.66} & \textbf{0.75} & \textbf{0.81} \\  
                            & $Deeplabv3+_F$           & 0.75          & 0.7           & 0.72          & 0.72          & 0.77          & 0.69          & 0.65          & 0.71          & 0.59          \\  
                            & $Deeplabv3+_P$           & \textbf{0.78} & \textbf{0.81} & \textbf{0.76} & \textbf{0.86} & \textbf{0.82} & \textbf{0.84} & \textbf{0.8}  & \textbf{0.88} & \textbf{0.79} \\ \hline
\multirow{2}{*}{3D}         & $U-net_F$                & 0.67          & 0.72          & 0.68          & 0.65          & 0.61          & 0.65          & 0.59          & 0.64          & 0.62          \\ 
                            & $U-net_P$                & \textbf{0.74} & \textbf{0.76} & \textbf{0.65} & \textbf{0.71} & \textbf{0.7}  & \textbf{0.66} & \textbf{0.68} & \textbf{0.72} & \textbf{0.71} \\ \hline
\end{tabular}
\end{table*}

Till today, developing semantic segmentation models are in a continuous active research, moreover this research took into account solely three models, namely, Deeplabv3+, U-net, and Segnet, respectively, and their performance with more attention to the input size, whether we fed the models the full image or some patches of the image.
The decided image size for all devices is set to be 572x572.
A famous denoising algorithm, namely, block-matching 3D filtering (BM3D), is applied to denoise the resized images \cite{dabov2007image}.
The choice of patch size is based on best reported results in \cite{deep1alsaih}, 
which is 128x128, knowing that, in the same reference, many patch sizes were tested.
To conclude, the sizes of the input to the deep models are as following:
\begin{itemize}
    \item 2D full image size: 572x572x1
    \item 2.5 and 3D full image size is: 384x384xN*, where N is = 3 if the image is in 2.5D, Or more for 3D.
    \item 2D patch size: 128x128x1
    \item 2.5D patch size: 128x128x3
    \item 3D patch size: 128x128xN*, where N depends on the volume size, such as Cirrus has 128 images for one volume (one eye), hence the size of the volume would be 128x128x128 and so on.
\end{itemize}

The architecture adjustments are described as follows:

\begin{enumerate}
\item\textbf{U-net \cite{ronneberger2015u}}: the primary focus of this network was on medical imaging applications.
The model of U-net is a type of encoder-decoder architecture, which employed the use of patches to solve the lack of ground truth data.
The architecture comprehends five sequential blocks, namely, convolution layers, ReLU layers, and Maxpooling layers, and was then followed by five additional blocks of upsampling layers.
The initial component represented by the Maxpooling, reduces the dimensionality of feature maps and captures high-level information, whereas the subsequent part regains the spatial resolution.
This paper considers padding operations to ensure that the input image equals the size of the convoluted or resulted image.

\item\textbf{Segnet \cite{badrinarayanan2017segnet}}: the model consists of a 5-block encoder and a 5-block decoder, utilising a pre-trained VGG16 network \cite{simonyan2014very} before recovering spatial resolution.
Unlike the U-net, this approach saves solely the max-pooling indices, rather than conveying the entire feature map from the encoder to the decoder, which requires significant memory size.
In this research, VGG16 was not utilized; instead, the whole model was run from zero to hero, no transfer learning was also used.
Segnet blocks are designed to be equivalent in number to U-net blocks. 

\item\textbf{Deeplabv3+ \cite{chen2018encoder}}: is a model that in order to recover the spatial resolution, employs atrous convolutions operations and two types of upsampling layers.
The Deeplabv3+ model structure has applied the Xception architecture as the bedrock element to the Deeplabv3+ architecture.
Initially, the model has employed a 1x1 convolution and three 3x3 atrous convolutions with different rates such as 6, 12, and 18, which is named as Atrous Spatial Pyramid Pooling (ASPP).
Dilated convolution or atrous convolution operations introduce an additional variable to the vanilla convolution called the dilation rate. 
Dilation rate expresses the spacing separation of the values in a kernel.
\end{enumerate}

It is important to point out that 75\% is the fixed amount of overlap chosen to extract patches from images.
Then the images are augmented using (translation, rotation operations).
All training sets are trained with a popular optimizer called Adam with a decay rate of 0.95 as well as the moving average of the squared gradient.
Initializing the learning rate at a value of 0.001 to 0.0001 was employed to the models.
The training of each scanner is performed 
Each scanner dataset is trained individually.
100 epochs as well as shuffling after each epoch is utilized for each training set.
Balancing the weight of the four classes is crucial, thus weighted cross-entropy loss is applied to the study.
Furthermore, no external images from other resources were added to the training phase of this study.
The evaluation of each deep model applied the function of dice similarity coefficient score (DSC).

\begin{equation}
\label{eqn:dice}
DSC = \frac{2TP}{2TP + FP + FN}
\end{equation} 
TP represents the number of true positive pixels, FP denotes the false positive pixels, and FN indicates the false negative pixels.

\section{Results and Discussion}\label{sec:exp-res}

This study's data validation relies on the cross-validation algorithm of 3k-fold.
Training involved 48 volumes, with 16 volumes from each scanner, while testing used 22 volumes, consisting of 8 volumes from the Cirrus scanner and the Spectralis device and 6 volumes from the Topcon scanner.
Consequently, models were trained using B-scans that contained no less than one pixel, which belongs to fluid’s class, resulting in ameliorated performance on two dataset, namely, Cirrus and Spectralis.
Models that contained only unhealthy B-scans trained on the Topcon dataset failed to enhance the model's performance.
The advantage of this training procedure lies in its ability to shorten the training period, in addition to preventing the inclusion of redundant images that are composed primarily of background pixels.
One can notice in most cases in Table~\ref{tab:my-table11}, a superiority in patch-based system’s performance.
The principal benefit of employing patches is their capacity to capture images at various scales.
Observing various neighboring pixels at the same ROI results in improved spatial consistency.
The resultant image covered some holes when using overlapping patches, which were supposed to be uniform.
However, all holes were not covered, plus an additional closing function was performed to cover the remaining cavities.
Patched 2D models perform better than 2.5D models on Cirrus scanner volumes only, and not in Spectralis or Topcon volumes.
3D models did not outperform the 2D or 2.5 models, that might be due to the lack of 3 data.
SRF fluid is the most recognized between all retinal cysts.
Hence, we strongly recommend using patches rather than full images due to their power in enhancing the performance and their ability to work as a denoiser element.
To the best of our knowledge, patch-based models have not been adequately analyzed for detecting and segmenting retinal fluids in various dimensions like in 2D, 2.5D, and 3D. 

\section{Conclusion}\label{sec:con}

In this work, we studied, analyzed and modified the 3D deep model in \cite{alsaih2020retinal2} to match the methods used in \cite{deep1alsaih} and \cite{alsaih2020retinal1} in order to make a fair comparison between the dimensions.
The retinal fluids were better identified when patches were fitted to the deep networks rather than the full image.
Overlapping patches work as a denoiser, maybe because they provide the networks with more images about the same ROI pixels at different locations and sizes.
2D and 2.5D models outperformed the 3D models, and SFR fluid was the most recognized by the models.
3D models could be enhanced if we have more data, in which we believe it could outperform all other models.
The reason could be, depth information might make more meaningful sense than fitting only the width information.
For future work, we recommend using patches and maybe fuse different sizes of patches which can track small cysts, which they sometimes were missed or mislocalized.


\bibliographystyle{IEEEtran}
\bibliography{IEEEabrv,references}

\begin{thebibliography}{10}
\providecommand{\url}[1]{#1}
\csname url@samestyle\endcsname
\providecommand{\newblock}{\relax}
\providecommand{\bibinfo}[2]{#2}
\providecommand{\BIBentrySTDinterwordspacing}{\spaceskip=0pt\relax}
\providecommand{\BIBentryALTinterwordstretchfactor}{4}
\providecommand{\BIBentryALTinterwordspacing}{\spaceskip=\fontdimen2\font plus
\BIBentryALTinterwordstretchfactor\fontdimen3\font minus
  \fontdimen4\font\relax}
\providecommand{\BIBforeignlanguage}[2]{{%
\expandafter\ifx\csname l@#1\endcsname\relax
\typeout{** WARNING: IEEEtran.bst: No hyphenation pattern has been}%
\typeout{** loaded for the language `#1'. Using the pattern for}%
\typeout{** the default language instead.}%
\else
\language=\csname l@#1\endcsname
\fi
#2}}
\providecommand{\BIBdecl}{\relax}
\BIBdecl

\bibitem{fung2007optical}
A.~E. Fung, G.~A. Lalwani, P.~J. Rosenfeld, S.~R. Dubovy, S.~Michels, W.~J.
  Feuer, C.~A. Puliafito, J.~L. Davis, H.~W. Flynn~Jr, and M.~Esquiabro, ``An
  optical coherence tomography-guided, variable dosing regimen with
  intravitreal ranibizumab (lucentis) for neovascular age-related macular
  degeneration,'' \emph{American journal of ophthalmology}, vol. 143, no.~4,
  pp. 566--583, 2007.

\bibitem{alsaih2017machine}
K.~Alsaih, G.~Lemaitre, M.~Rastgoo, J.~Massich, D.~Sidib{\'e}, and
  F.~Meriaudeau, ``Machine learning techniques for diabetic macular edema (dme)
  classification on sd-oct images,'' \emph{Biomedical engineering online},
  vol.~16, no.~1, p.~68, 2017.

\bibitem{girish2018segmentation}
G.~Girish, B.~Thakur, S.~R. Chowdhury, A.~R. Kothari, and J.~Rajan,
  ``Segmentation of intra-retinal cysts from optical coherence tomography
  images using a fully convolutional neural network model,'' \emph{IEEE journal
  of biomedical and health informatics}, vol.~23, no.~1, pp. 296--304, 2018.

\bibitem{abdolrezadeep2019re}
A.~Rashno, D.~Koozekanani, and K.~Parhi, ``Detection and segmentation of
  various types of fluids with graph shortest path and deep learning
  approaches,'' in \emph{MICCAI Retinal OCT Fluid Challenge (RETOUCH)}, Sep.
  2017.

\bibitem{lu2017retinal}
D.~Lu, M.~Heisler, S.~Lee, G.~Ding, M.~V. Sarunic, and M.~F. Beg, ``Retinal
  fluid segmentation and detection in optical coherence tomography images using
  fully convolutional neural network,'' in \emph{MICCAI Retinal OCT Fluid
  Challenge (RETOUCH)}, 2017.

\bibitem{yadav2017generalized}
S.~Yadav, K.~Gopinath, and J.~Sivaswamy, ``A generalized motion pattern and fcn
  based approach for retinal fluid detection and segmentation,'' in
  \emph{MICCAI Retinal OCT Fluid Challenge (RETOUCH)}, Sep. 2017.

\bibitem{apostolopoulos2017pathological2}
S.~Apostolopoulos, C.~Ciller, R.~Sznitman, and S.~De~zanet, ``Simultaneous
  classification and segmentation of cysts in retinal oct,'' in \emph{MICCAI
  Retinal OCT Fluid Challenge (RETOUCH)}, Sep. 2017.

\bibitem{sungdeep2019re}
S.~H. Kang, H.~S. Park, J.~Jang, and K.~Jeon, ``Deep neural networks for the
  detection and segmentation of the retinal fluid in oct images,'' in
  \emph{MICCAI Retinal OCT Fluid Challenge (RETOUCH)}, Sep. 2017.

\bibitem{chendeepdeep2019}
Q.~Chen, Z.~Ji, T.~Wang, Y.~Tand, C.~Yu, O.~I. Paul, and L.~B. Sappa,
  ``Automatic segmentation of fluid-associated abnormalities and pigment
  epithelial detachment in retinal sd-oct images,'' in \emph{MICCAI Retinal OCT
  Fluid Challenge (RETOUCH)}, Sep. 2017.

\bibitem{morley2017simultaneous}
D.~Morley, H.~Foroosh, S.~Shaikh, and U.~Bagci, ``Simultaneous detection and
  quantification of retinal fluid with deep learning,'' in \emph{MICCAI Retinal
  OCT Fluid Challenge (RETOUCH)}, Sep. 2017.

\bibitem{ruwandeep2019re}
R.~Tennakoon, A.~k. Gostar, R.~Hoseinnezhad, and A.~Bab-Hadiashar, ``Retinal
  fluid segmentation and classification in oct images using adversarial loss
  based cnn,'' in \emph{MICCAI Retinal OCT Fluid Challenge (RETOUCH)}, Sep.
  2017.

\bibitem{alsaih2020performance}
K.~Alsaih, M.~Z. Yusoff, T.~B. Tang, I.~Faye, and F.~M{\'e}riaudeau,
  ``Performance evaluation of convolutions and atrous convolutions in deep
  networks for retinal disease segmentation on optical coherence tomography
  volumes,'' in \emph{2020 42nd Annual International Conference of the IEEE
  Engineering in Medicine \& Biology Society (EMBC)}.\hskip 1em plus 0.5em
  minus 0.4em\relax IEEE, 2020, pp. 1863--1866.

\bibitem{oh2024gcn}
D.~Oh, J.~Moon, K.~Park, W.~Kim, S.~Yoo, H.~Lee, and J.~Yoo, ``Gcn-assisted
  attention-guided unet for automated retinal oct segmentation,'' \emph{Expert
  Systems with Applications}, vol. 249, p. 123620, 2024.

\bibitem{pavani2024robust}
P.~G. Pavani, B.~Biswal, T.~K. Gandhi, and A.~R. Kota, ``Robust semantic
  segmentation of retinal fluids from sd-oct images using fam-u-net,''
  \emph{Biomedical Signal Processing and Control}, vol.~87, p. 105481, 2024.

\bibitem{deep1alsaih}
K.~Alsaih, M.~Z. Yusoff, T.~B. Tang, I.~Faye, and F.~M{\'e}riaudeau, ``Deep
  learning architectures analysis for age-related macular degeneration
  segmentation on optical coherence tomography scans,'' \emph{Computer methods
  and programs in biomedicine}, vol. 195, p. 105566, 2020.

\bibitem{alsaih2020retinal1}
K.~Alsaih, M.~Z. Yusoff, I.~Faye, T.~B. Tang, and F.~Meriaudeau, ``Retinal
  fluid segmentation using ensembled 2-dimensionally and 2.5-dimensionally deep
  learning networks,'' \emph{IEEE Access}, vol.~8, pp. 152\,452--152\,464,
  2020.

\bibitem{alsaih2020retinal2}
K.~Alsaih, M.~Yusoff, T.~Tang, I.~Faye, and F.~M{\'e}riaudeau, ``Retinal fluids
  segmentation using volumetric deep neural networks on optical coherence
  tomography scans,'' in \emph{2020 10th IEEE International Conference on
  Control System, Computing and Engineering (ICCSCE)}.\hskip 1em plus 0.5em
  minus 0.4em\relax IEEE, 2020, pp. 68--72.

\bibitem{bogunovic2019retouch}
H.~Bogunovi{\'c}, F.~Venhuizen, S.~Klimscha, S.~Apostolopoulos,
  A.~Bab-Hadiashar, U.~Bagci, M.~F. Beg, L.~Bekalo, Q.~Chen, C.~Ciller
  \emph{et~al.}, ``Retouch: The retinal oct fluid detection and segmentation
  benchmark and challenge,'' \emph{IEEE transactions on medical imaging},
  vol.~38, no.~8, pp. 1858--1874, 2019.

\bibitem{dabov2007image}
K.~Dabov, A.~Foi, V.~Katkovnik, and K.~Egiazarian, ``Image denoising by sparse
  3-d transform-domain collaborative filtering,'' \emph{Image Processing, IEEE
  Transactions on}, vol.~16, no.~8, pp. 2080--2095, 2007.

\bibitem{ronneberger2015u}
O.~Ronneberger, P.~Fischer, and T.~Brox, ``U-net: Convolutional networks for
  biomedical image segmentation,'' in \emph{International Conference on Medical
  image computing and computer-assisted intervention}.\hskip 1em plus 0.5em
  minus 0.4em\relax Springer, 2015, pp. 234--241.

\bibitem{badrinarayanan2017segnet}
V.~Badrinarayanan, A.~Kendall, and R.~Cipolla, ``Segnet: A deep convolutional
  encoder-decoder architecture for image segmentation,'' \emph{IEEE
  transactions on pattern analysis and machine intelligence}, vol.~39, no.~12,
  pp. 2481--2495, 2017.

\bibitem{simonyan2014very}
K.~Simonyan and A.~Zisserman, ``Very deep convolutional networks for
  large-scale image recognition,'' \emph{arXiv preprint arXiv:1409.1556}, 2014.

\bibitem{chen2018encoder}
L.-C. Chen, Y.~Zhu, G.~Papandreou, F.~Schroff, and H.~Adam, ``Encoder-decoder
  with atrous separable convolution for semantic image segmentation,'' in
  \emph{Proceedings of the European conference on computer vision (ECCV)},
  2018, pp. 801--818.

\end{thebibliography}

\end{document}